\documentclass[proof]{pasj00}
\draft

\begin{document}
\SetRunningHead{S. F. Yamada et al.}{Possibility of Gravitational Lensing for High-z Quasars}
\Received{2003 February 28}
\Accepted{2003 June 13}

\title{Are Two $z \sim 6$ Quasars Gravitationally Lensed ?\altaffilmark{*}}

\author{
        Sanae F. \textsc{Yamada},      \altaffilmark{1}
        Yasuhiro \textsc{Shioya},      \altaffilmark{1}
        Yoshiaki  \textsc{Taniguchi},  \altaffilmark{1}
        Takashi  \textsc{Murayama},    \altaffilmark{1}\\
        Masaru  \textsc{Ajiki},        \altaffilmark{1}
        Tohru  \textsc{Nagao},         \altaffilmark{1}
        Shinobu S.  \textsc{Fujita},   \altaffilmark{1}
        Kazuyoshi  \textsc{Umeda},     \altaffilmark{1}\\
        Yutaka  \textsc{Komiyama},     \altaffilmark{2}
        Hiroshi \textsc{Karoji},       \altaffilmark{2}
        Hiroyasu \textsc{Ando},        \altaffilmark{3}
        Masanori  \textsc{Iye},        \altaffilmark{3}\\
        Nobunari  \textsc{Kashikawa},  \altaffilmark{3} and
        Keiichi  \textsc{Kodaira},     \altaffilmark{4}
        }

\altaffiltext{1}{Astronomical Institute, Graduate School of Science,
        Tohoku University, Aramaki, Aoba,\\Sendai 980-8578}
\email{fiddle@astr.tohoku.ac.jp}
\altaffiltext{2}{Subaru Telescope, National Astronomical Observatory of Japan,\\
        650 N.A'ohoku Place, Hilo, HI 96720, USA}
\altaffiltext{3}{National Astronomical Observatory,
        2-21-1 Osawa, Mitaka, Tokyo 181-8588}
\altaffiltext{4}{The Graduate University for Advanced Studies (SOKENDAI),\\
        Shonan Village, Hayama, Kanagawa 240-0193}


\maketitle

\footnotetext[*]{Based on data collected at
        Subaru Telescope, which is operated by
        the National Astronomical Observatory of Japan.}

\begin{abstract}
Several high-$z$ ($z > 5.7$) quasars have been found
during the course of Sloan Digital Sky Survey.
The presence of such very high-$z$ quasars is expected to put constraints on early structure formation.
On one hand, it is suggested that these most luminous objects at high redshift
are biased toward highly magnified objects by gravitational lensing.
To clarify the effect of gravitational lensing on high-$z$ quasars,
we began an imaging survey of intervening lensing galaxies.
Indeed, our previous optical image showed that SDSSp J104433.04$+$012502.2 at $z$=5.74 is gravitationally magnified by a factor of 2.
In this paper, we report our new optical imaging of two other high-$z$ quasars, SDSSp J103027.10$+$052455.0 at $z$=6.28 and SDSSp J130608.26$+$035626.3 at $z$=5.99. Since we found neither an intervening galaxy nor a counter image with $i^{\prime} < 25.4-25.8$ around each quasar, we conclude that they are not strongly magnified regardless that a lens galaxy is dusty.

\end{abstract}

\section{Introduction}

The Sloan Digital Sky Survey (SDSS: e.g., York et al. 2000)
has been producing the discovery of high-redshift quasars at $z \approx 6$ (Fan et al. 2000, 2001, 2003);
the most distant one known to date is SDSSp J114816.64+525150.2
at $z=6.43$ (Fan et al. 2003).
One significant feature seems to be that all of these $z \sim 6$ SDSS quasars
are exceptionally bright.
As claimed by Wyithe \& Loeb (2002b; hereafter WL02),
the presence of such very bright quasars at $z \sim 6$ raises
the following two problems: (1) If the central engine of these
quasars is an accreting super-massive black hole, and shines
near the Eddington accretion rate, a super-massive black hole
with mass exceeding $\sim 3 \times 10^9 M_\odot$ was already formed
beyond $z=6$. This challenges any theoretical models for early structure formation
(Turner 1991; Haiman, Loeb 2001). (2) Also, the absolute luminosities
of the $z \sim 6$ quasars are systematically higher than
expected from the SDSS survey criteria (Fan et al. 2000, 2001),
suggesting that the luminosity function of the $z \sim 6$ quasars
is significantly biased to higher luminosities.
These two problems led WL02 to propose that a third of the known $z \sim 6$ quasars
may be magnified by a factor of 10 or more by intervening gravitational lenses, although the lensing probability depends on the slope of the luminosity function (Wyithe, Loeb 2002a). Comerford et al. (2002) also estimated the probability of gravitational lensing magnification by a factor of 10 or more for the four SDSS quasars at $z \sim 6$, to be $\sim$ 100\% given that the LF of the quasar is steep or extends to faint magnitudes.

During the course of our optical deep-survey program on
Ly$\alpha$ emitters at $z \approx 5.7$ in the field surrounding
the quasar SDSSp J104433.04$-$012502.2 at $z=5.74$ (Ajiki et al. 2002; Taniguchi et al. 2003),
we found that a faint galaxy with $m_B$(AB) $\approx 25$ is located at
\timeform{1".9} southwest of this quasar (Shioya et al. 2002).
We estimated the redshift of this galaxy to be $z \sim 2.1$, and
concluded that SDSSp J104433.04$-$012502.2 could be magnified by a factor of two.
Although this fact supports the suggestion by WL02, further information on other very high-$z$ quasar is necessary.
In this paper, we present our new results of a lensing galaxy survey
around two SDSS quasars, SDSSp J103027.10$+$052455.0 at $z = 6.28$ and
SDSSp J130608.26$+$035626.3 at $z = 5.99$.
Throughout this paper,
we adopt a flat universe with $\Omega_{\rm m} = 0.3$,
$\Omega_{\Lambda} = 0.7$,
and $h=0.7$, where $h = H_0/($100 km s$^{-1}$ Mpc$^{-1}$)
and we use magnitudes given in the AB system.

\section{Observations and Results}

\subsection{Observations}

Deep optical imaging observations were made with the Suprime-Cam
(Miyazaki et al. 2002) on
the 8.2 m Subaru Telescope (Kaifu 1998) at Mauna Kea Observatories.
The Suprime-Cam consists of ten 2k$\times$4k CCD chips and
provides a very wide field of view: $34^\prime \times 27^\prime$
with a \timeform{0".2}/pixel resolution.
Using this facility, we made a very deep $i^{\prime}$-band imaging survey
in the fields around 
SDSSp J103027.10$+$052455.0 at a redshift of 6.28 and 
SDSSp J130608.26$+$035626.3 at a redshift of 5.99
(Fan et al. 2001; Becker et al. 2001).
A summary of the imaging observations
is given in table 1. All of the observations were made under
photometric conditions, and the seeing size was between \timeform{1".2}
and \timeform{1".8} during the run.

The CCD data were reduced and
combined using IRAF and mosaic-CCD data-reduction software
developed by Yagi et al. (2002).
Since the observation was made during the course of the Subaru Deep Field project (e.g., Kodaira et al. 2003), the available observing time for the purpose was limited, and thus we could not take photometric-standard star data. Therefore, in order to  calibrate our data, we used the SDSS Date Release 1 (DR1) catalog. We selected all point sources within 5 arcmin from each quasar from the SDSS DR1 catalog. In this procedure, we selected the point sources with $18 < i^{\prime} < 20$, because objects with $i^{\prime} < 18$ are too bright in our images, while objects with $i^{\prime} > 20$ are too faint to be well calibrated in the SDSS DR1 catalog.

After the calibration, we obtained the $i^\prime$ magnitudes of each quasar: 
$i^{\prime}=23.23{\pm}0.07$ for SDSSp J103027.10$+$052455.0,  and
$i^{\prime}=21.95{\pm}0.03$ for SDSSp J130608.26$+$035626.3.
The magnitude of SDSSp J103027.10$+$052455.0 is consistent with that of Fan et al. (2001). On the other hand, the magnitude of SDSSp J130608.26$+$035626.3 is 0.63 mag brighter than that of Fan et al. (2001). The difference may be because our image of SDSSp J130608.26$+$035626.3 has a deeper limiting magnitude, and thus our photometric accuracy is better than that of the SDSS data for faint objects.\\

\subsection{SDSSp J103027.10$+$052455.0}

The $i^{\prime}$-band images around SDSSp J103027.10$+$052455.0 are shown in the left panel of figure 1. We did not detect any galaxy within 6 arcsec from the quasar SDSSp J103027.10$+$052455.0.  No possible counter image of the quasar was identified either.
We checked the radial profile of the quasar in this image, and found that the image is consistent with that of point sources. Therefore, we conclude that the quasar may not be gravitationally magnified by a foreground galaxy.

Fan et al. (2003) analyzed high-resolution images of SDSSp J103027.10$+$052455.0
observed by the Advanced Camera for Surveys on the Hubble Space Telescope
and the Near Infrared Camera on the Keck I telescope, and found that the
quasar appears to be an unresolved point source. They also mentioned that there are no close companions with $K^{\prime} < 21$ within $10^{\prime \prime}$ from the quasar. These results are consistent with our results. Schwartz (2002) pointed out that SDSSp J103027.10$+$052455.0 is not consistent with a point source from the X-ray observation by the Chandra X-ray Observatory, and interpreted that the quasar may be gravitationally lensed. However, we found neither a galaxy nor a counter image within 6 arcsec from the quasar. This fact shows that the quasar is not strongly magnified.

\subsection{SDSSp J130608.26$+$035626.3}

The $i^{\prime}$-band image around SDSSp J130608.26$+$035626.3 is shown in the right panel of figure 1. Again, we found neither a galaxy within 3 arcsec from the quasar nor a counter image of the quasar within 5 arcsec around SDSSp J130608.26$+$035626.3.
We also checked the radial profile of the quasar in this image, and found that the image is consistent with that of point sources.

Fan et al. (2003) also analyzed the high-resolution images of SDSSp J130608.26$+$035626.3 derived from the Astrophysical Research Consortium 3.5 m telescope, and concluded that the quasar also appears to be an unresolved point source.
Ivanov (2002) obtained a deep $I$-band image of the quasar from the VLT with FORS1, and showed the radial profile of SDSSp J130608.26$+$035626.3 is consistent with that of a point source. Our result concluded that the quasar is an unresolved point source as well. Schwartz (2002) also pointed out that the quasar is consistent with a point source.
Becker et al. (2001) found a candidate of damped Ly$\alpha$ systems
at $z = 4.86$ in the spectrum of the quasar.
However, we found neither an optical counterpart nor a counter image of the system within 3 arcsec from the quasar in our $i^{\prime}$-band image. We conclude that the gravitational magnification, if any, of the quasar is not strong.

\section{Discussion}

We cannot find any lens galaxy in our $i^{\prime}$ images around SDSSp J103027.10$+$052455.0 at $z$=6.28 and SDSSp J130608.26$+$035626.3 at $z$=5.99.
However, these results do not mean that the quasars are not magnified by gravitational
lensing of intervening galaxies, since galaxies fainter than our limiting magnitudes may still be present and
magnify the light of quasars.
We, therefore, evaluated the maximum magnification factor by galaxies fainter
than the limiting magnitude.
In order to estimate how the quasar is
gravitationally magnified, we need information about both the redshift
and the stellar velocity dispersion of the lens galaxy.
Although there is no information on $z$ for a possible lens galaxy,
we can estimate the maximum luminosity of galaxies which is consistent with
our observations.
Second, we also estimated the maximum velocity dispersion of the dark matter halo of galaxies using the Tully--Fisher relation.
Then, we estimated the maximum magnification factor based on the singular isothermal sphere model for gravitational lensing.

\subsection{Absolute Magnitude of the Lens Galaxy}

The maximum luminosity, which is consistent with our observation,
depends on the shape of the spectral energy distribution (SED) of a lens galaxy.
The SED is
mainly determined by the following four factors:
(1) the radiation from stars in the galaxy,
(2) the extinction by dust in the galaxy, itself,
(3) the redshift of the galaxy, and
(4) the absorption by intergalactic neutral hydrogen between
the galaxy and us.
We treat the above factors in the following way.

We use the population synthesis model GISSEL96,
which is a revised version of Bruzual and Charlot (1993)
(see also Leitherer et al. 1996), to calculate an SED from stellar components.
The SED of the galaxy is then determined by its star-formation history.
The SEDs of local galaxies are well reproduced
by models whose star-formation rate
declines exponentially ($\tau$ model); i.e.,
$SFR(t) \propto \exp (-t/\tau)$,
where $t$ is the age of the galaxy and $\tau$ is the time scale of star formation.
It is noted that different combinations between $t$ and $\tau$ generate a similar shape of SED in some cases.
In this work,
we therefore used $\tau=1$ Gyr models with Salpeter's initial mass function
(the power index of $x=1.35$ and the stellar mass range of
$0.01 \leq m/M_\odot \leq 125$) to derive various SED types.
We adopted the solar metallicity, $Z=0.02$.
For these models, we calculated SEDs with ages of
$t$ = 0.1, 0.5, 1, 2, 3, 4, and 8 Gyr, and
called them SED1, SED2, SED3, SED4, SED5, SED6, and SED7, respectively;
note that the SED templates derived by Coleman et al. (1980),
elliptical galaxies (the bulges of M 31 and M 81), Sbc, Scd, and Irr
correspond to those of SED7 ($t=8$ Gyr),
SED6 ($t=4$ Gyr), SED5 ($t=3$ Gyr), and SED4 ($t=2$ Gyr), respectively.
We did not use an SED whose $t$ was larger than the age of the universe at $z$.
For our adopted model of the universe,
the age of the universe exceeds 1, 2, 3, 4, and 8 Gyr at $z=5.7$, 3.2, 2.2, 1.6, and 0.6, respectively.
For simplicity, we assumed that the dust extinction of the lensing galaxy to be $A_V=0$.
The effect of $A_V$ on our discussion is mentioned below.
As for the absorption by intergalactic neutral hydrogen,
we used the average optical depth derived by Madau et al. (1996).
In this procedure, we adopted an allowed redshift of between $z=0$ and
those of quasars with a redshift bin of $\Delta z$ = 0.1.
Figure 2 shows the $B$-band absolute magnitude of galaxies
corresponding to the observed limiting $i^{\prime}$-band magnitude as a function of $z$ and SED.

\subsection{Stellar Velocity Dispersion of the Lens Galaxy}

To estimate of the magnification factor by gravitational lensing, 
it is necessary to know the stellar velocity dispersion
($\sigma_v$) of the lens galaxy.
Given the redshift of a lens galaxy, we could estimate its absolute
magnitude and then estimate a probable value of the stellar velocity
dispersion. However, since there is no information about $z$, we tried to estimate the stellar velocity dispersion
as a function of both the redshift and SED assuming that the Tully--Fisher relation
established for a sample of galaxies in the local universe is also valid at a high redshift.

Here, we used the Tully--Fisher relation derived by Sakai et al. (2000),

\begin{equation}
M_B  =  -8.07 (\log W_{20} - 2.5) - 19.88,
\end{equation}
where $M_B$ is the absolute $B$ magnitude and
$W_{20}$ is the full width at 20\% of the maximum velocity.
Assuming that $\sigma_v = V_{\rm rot}/\sqrt{2} = W_{20}/2\sqrt{2}$ where
$V_{\rm rot}$ is the rotation velocity,
we estimated $\sigma_v$ as a function of $z$ and SED.
We evaluated $M_B$ using the method given in subsection 3.1.
The results are shown in figure 3. In the case of SDSSp J103027.10$+$052455.0, $\sigma_v \lesssim$ 300 km s$^{-1}$ at $z< 5.5$.
In the case of SDSSp J130608.26$+$035626.3, $\sigma_v \lesssim$ 200 km s$^{-1}$ at $z< 5.5$.

\subsection{Magnification Factor by the Gravitational Lensing}

We are now ready to estimate the magnification factor by
gravitational lensing. We adopt the singular isothermal sphere
(SIS) model for simplicity (e.g., Binney, Merrifield 1998).
In this model, the magnification factor for the brighter source can be expressed as

\begin{equation}
M_+ = \frac{\theta}{\theta - \theta_{\rm E}},
\end{equation}
where $\theta$ is the angle between the lens galaxy and
the source (i.e., SDSSp J103027.10$+$052455.0 and SDSSp J130608.26$+$035626.3 in this case),
and $\theta_{\rm E}$ is the Einstein angle, defined as

\begin{equation}
\theta_{\rm E} = 4 \pi \left( \frac{\sigma_v}{c} \right)^2
\frac{D_{\rm LS}}{D_{\rm OS}},
\end{equation}
where $D_{\rm LS}$ is the angular diameter distance between the
lens galaxy and the source and $D_{\rm OS}$ is that between
the observer and the source. $M_+$ of the brighter image and $M_-$ of the counter image are related by
$M_+/M_- = M_+/(M_+-2)$.
In figure 4, we show
the magnification factor of the brighter source as a function of $z$ and the SED types. Here, we adopt $\theta = 1^{\prime \prime}$. The figures show that the maximum of $M_+$ is only $M_+ \sim 5$ (for SDSSp J103027.10$+$052455.0)
and $M_+ \sim 2$ (for SDSSp J130608.26$+$035626.3). On the other hand, $M_+$ is constrained from the condition that there is no counter image and results in  $M_+ < 2.31$ (SDSSp J103027.10$+$052455.0) and $M_+ < 2.11$ (SDSSp J130608.26$+$035626.3). The constraint that no lensing galaxy can be seen gives a similar result to that from the absence of any counter image.

In the case of $A_V > 0$, the maximum $M_+$, which is derived from the condition that there is no intervening galaxy, becomes larger. We emphasize, however, that the constraint on $M_+$ from the condition that there is no counter image is not changed. We therefore conclude that the two quasars are not strongly lensed. The gravitational lensing effect, if any, should be very modest.

\section{Concluding Remarks}
WL02 suggested that the SDSS $z \sim6$ quasars are systematically brighter
than expected. Namely, they investigated the probability that
a quasar with a rest-frame 1450 \AA ~ magnitude, $M_{1450}$(AB), and
a redshift of $z$ was selected into the survey using the survey selection
function given in Fan et al. (2001). They found that
the cumulative probability of $M_{1450}$(AB) for the $z \sim 6$ quasars
is different from the expected one at the 95\% significance level.
However, they also suggested that this inconsistence is removed if
the flux of one third of $z \sim 6$ quasars is overestimated by a
factor of 10.

In this paper, we cannot find any strong evidence that supports this suggestion.
We have found no candidate of a lensing galaxy brighter than
$i^{\prime} = 25.4$ around the $z$=6.28 quasar SDSSp J103027.10$+$052455.0 and
$i^{\prime} = 25.8$ around the $z$=5.99 quasar SDSSp J130608.26$+$035626.3.
We have also found no counter image of the two quasars.
We evaluated the possible magnification factor at the angle $\theta =1^{\prime \prime}$ using the SIS model:
1) $M_+ < 5$ for SDSSp J103027.10$+$052455.0, and
2) $M_+ < 2$ for SDSSp J130608.26$+$035626.3 in the case of $A_V=0$,
and to be large in the case of $A_V > 0$. However, the constraints on $M_+$ from the condition that there is no counter image shows that $M_+$ is not very large in any case of $A_V$. Therefore, we conclude that both quasars are not strongly gravitationally magnified.

Here, we mention a limitation on our analysis.
If the angle between a quasar and a lensing galaxy (and a counter image)
is much smaller than the seeing size, we cannot find evidence of 
gravitational lensing from our observations.
High-resolution deep imaging at wavelengths shorter than
the Lyman break of the quasar may give further constraints to gravitational lensing of this type.

It is known that the lensing optical depth increases with increasing
the redshift, and thus objects at higher redshift are more affected
by gravitational lensing from a statistical point of view
(e.g., Turner 1991; Barkana, Loeb 2000; WL02). Fan et al. (2003) noted that even without a detectable second image under high resolution, the possibility that luminous $z \sim 6$ quasars are modestly magnified by foreground galaxies is still high. 
As proposed by WL02, it will be very important to carefully search for lensing galaxies
towards all high-$z$ quasars by deeper
imaging with higher resolution to constrain the magnification factor.
To search for lensing galaxies that are overlapped with a quasar, high-resolution spectroscopy of the candidate of magnified quasars will also be important.

\bigskip

We would like to thank the Subaru Telescope staff
for their invaluable help.
We also thank T. Hayashino for his invaluable help.
This work was financially supported in part by
the Ministry of Education, Culture, Sports, Science and Technology
(Nos. 10044052 and 10304013). TN and MA are JSPS fellows.


\vspace{1cm}

\begin{figure}
\begin{center}
\begin{tabular}{cc}
\FigureFile(60mm,60mm){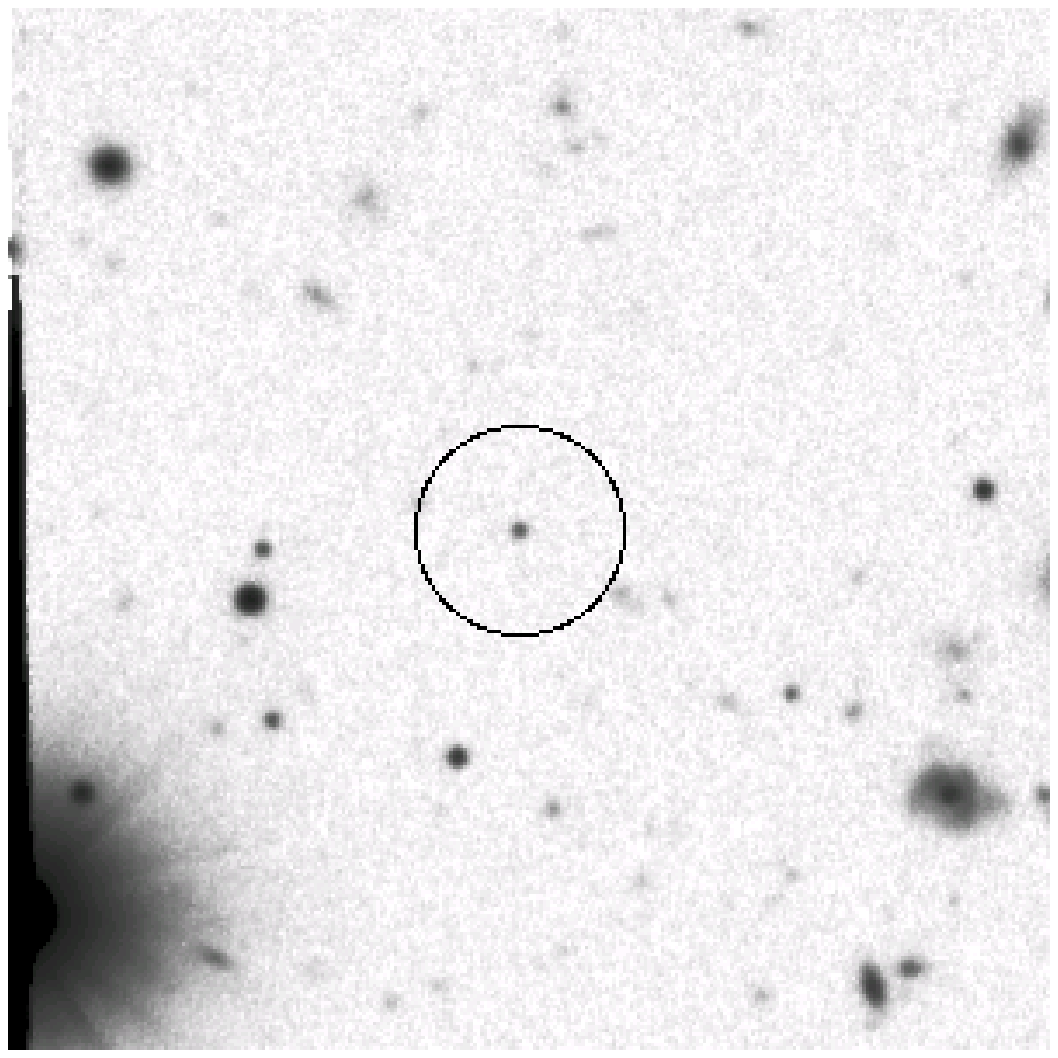}&
\FigureFile(60mm,60mm){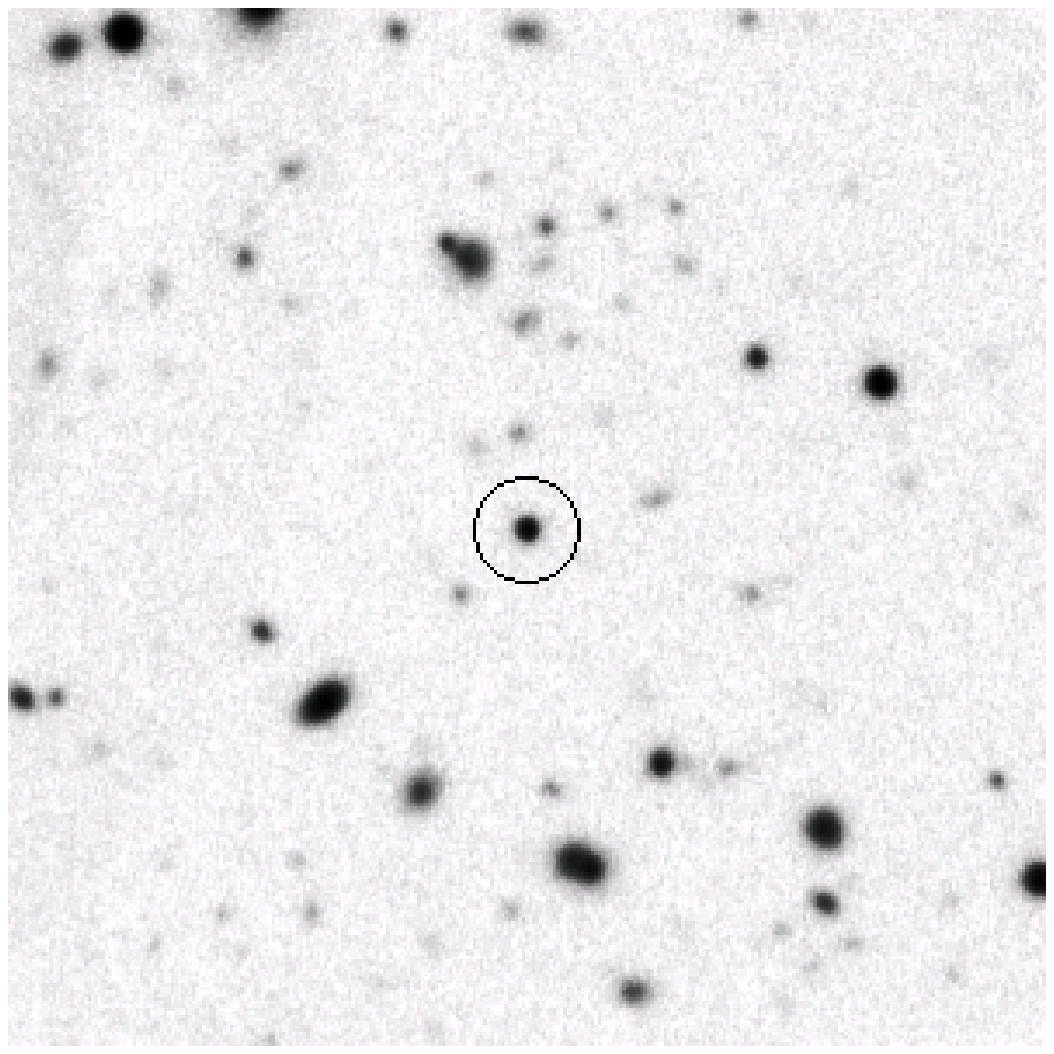}
\end{tabular}
\end{center}
\caption{
$i^{\prime}$-band images around SDSSp J103027.10+052455.0 (left) and
SDSSp J130608.26+035626.3 (right). The angular size of the circle in each panel
corresponds to $6^{\prime \prime}$ and $3^{\prime \prime}$.
No galaxy can be seen within the circle. }
\label{fig:fig1}
\end{figure}

\begin{figure}
\begin{center}
\begin{tabular}{cc}
\FigureFile(80mm,80mm){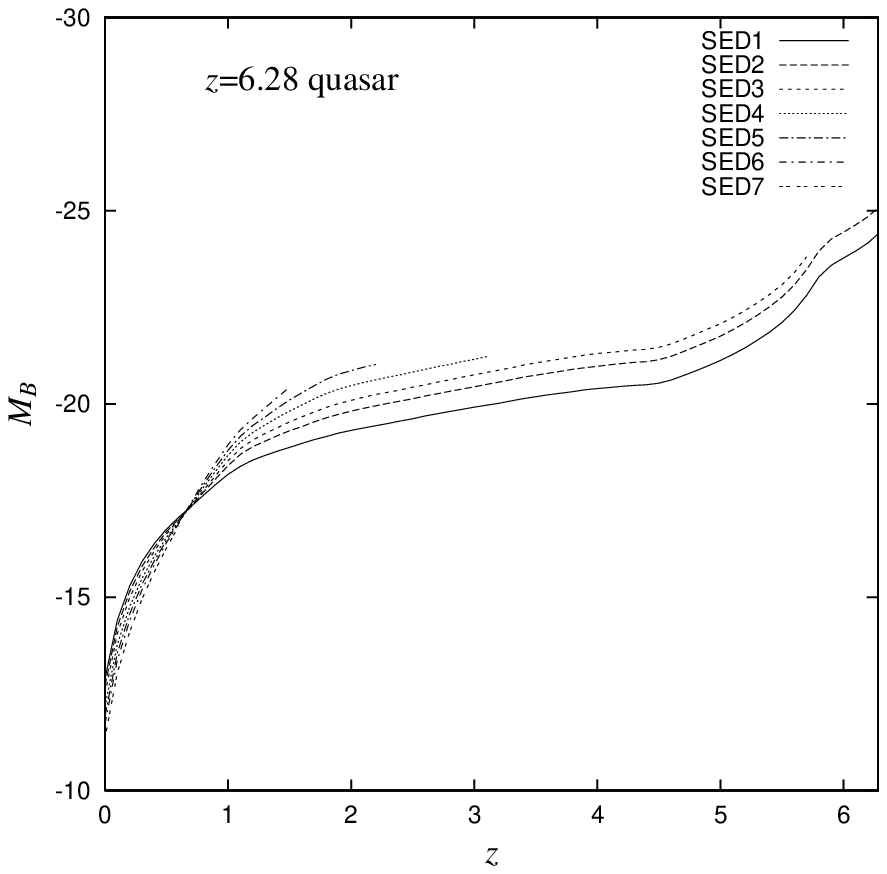}
\FigureFile(80mm,80mm){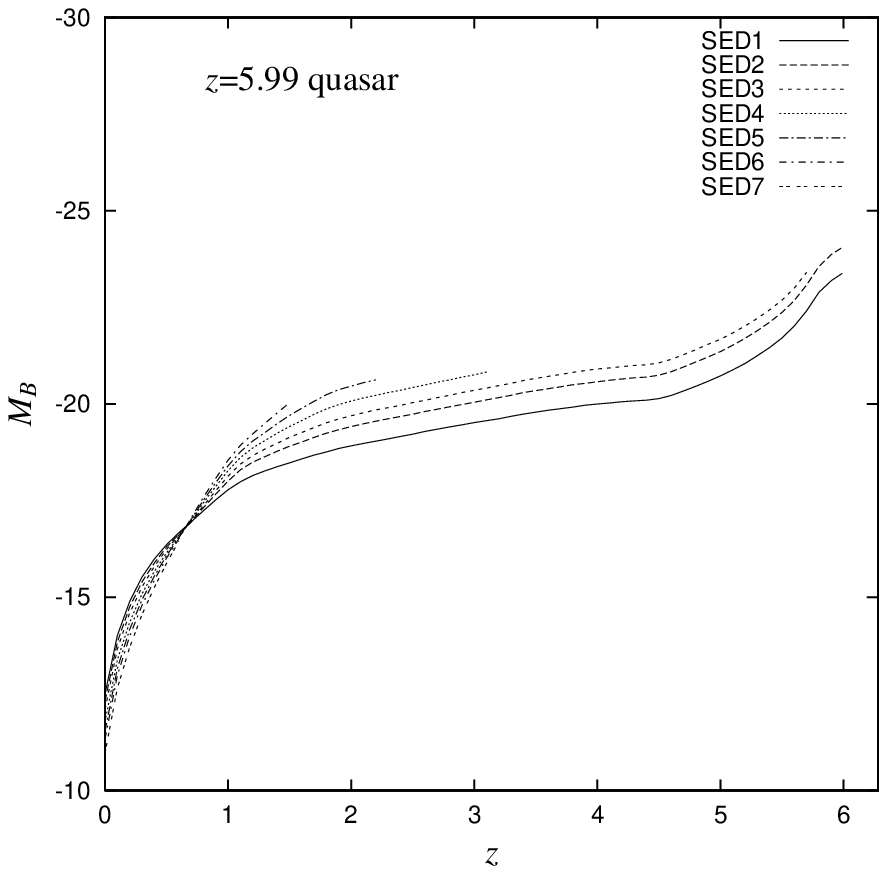}
\end{tabular}
\end{center}
\caption{
Distribution of the absolute magnitude of galaxies, which corresponds to
the limiting magnitude, as a function of the SED and the redshift.
}
\label{fig:fig2}
\end{figure}

\begin{figure}
\begin{center}
\begin{tabular}{cc}
\FigureFile(80mm,80mm){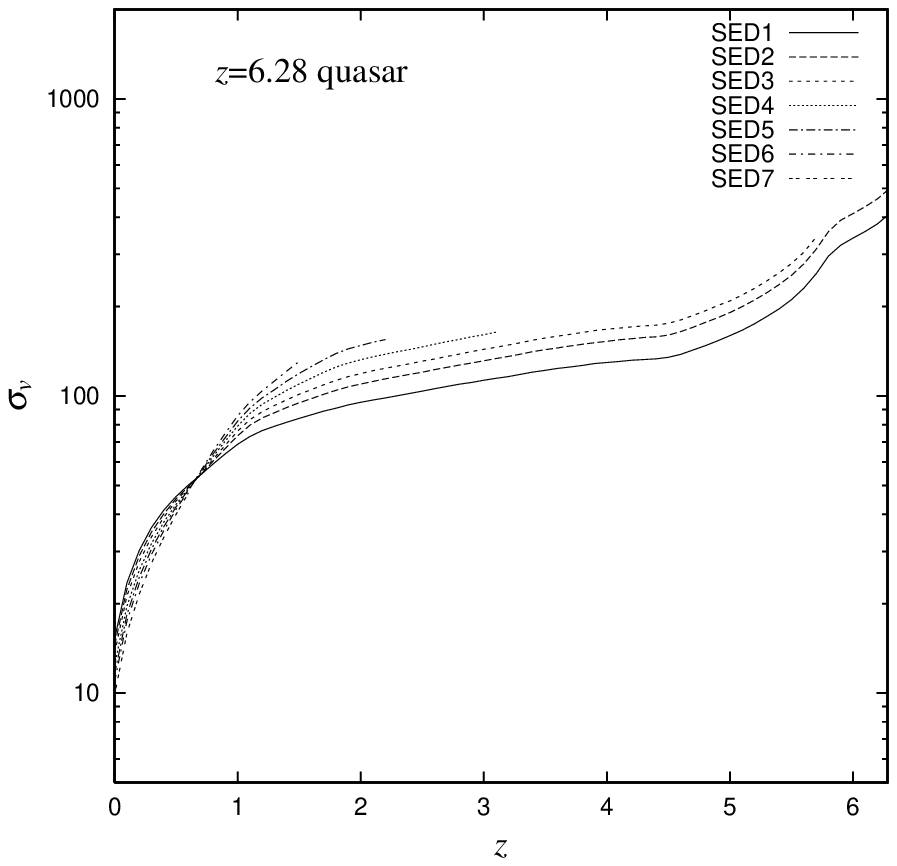}
\FigureFile(80mm,80mm){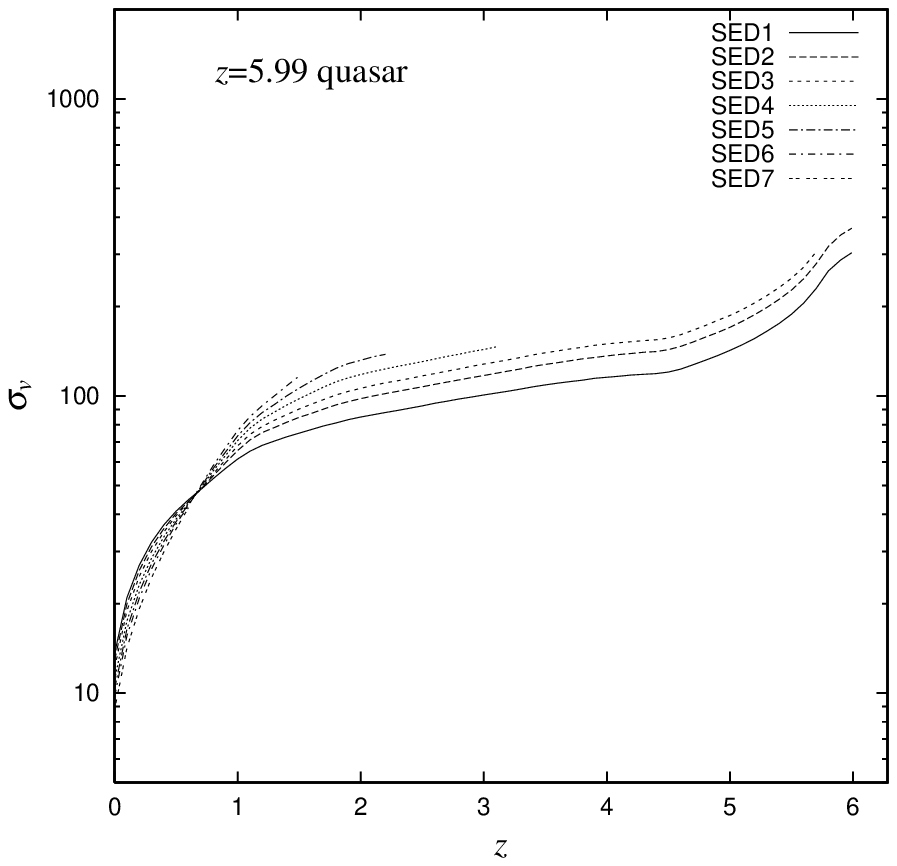}
\end{tabular}
\end{center}
\caption{
Distribution of the maximum velocity dispersion of the lens galaxy
with $> m_{\rm lim}$ as a function of the SED and the redshift.
}
\label{fig:fig3}
\end{figure}

\begin{figure}
\begin{center}
\begin{tabular}{cc}
\FigureFile(80mm,80mm){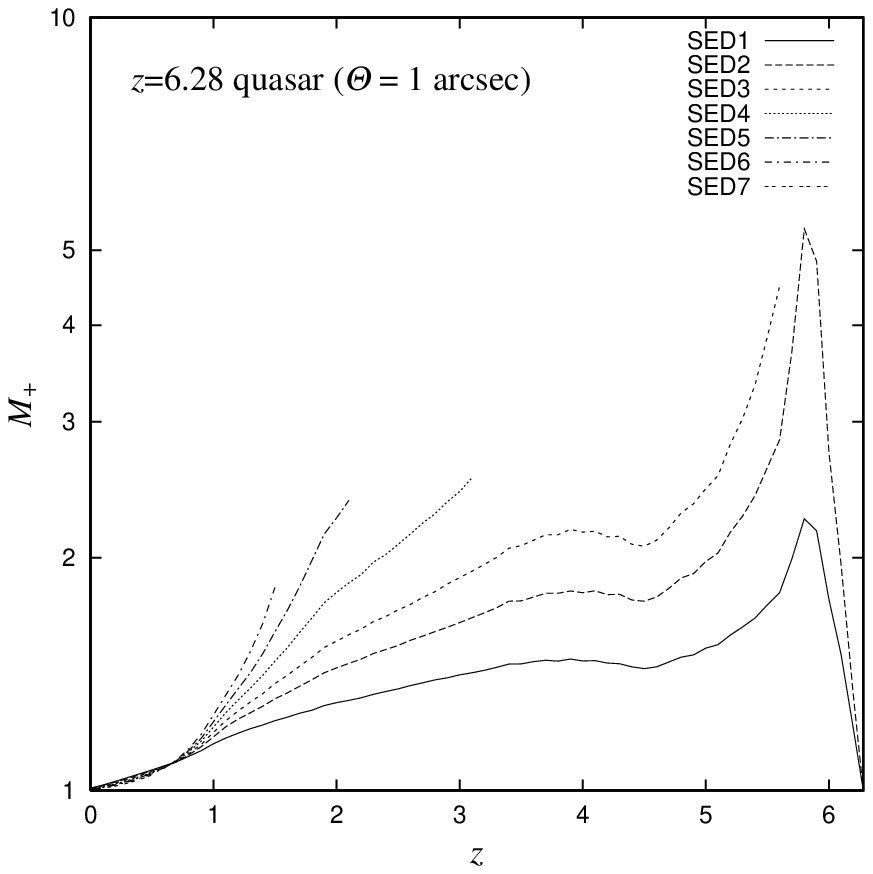}
\FigureFile(80mm,80mm){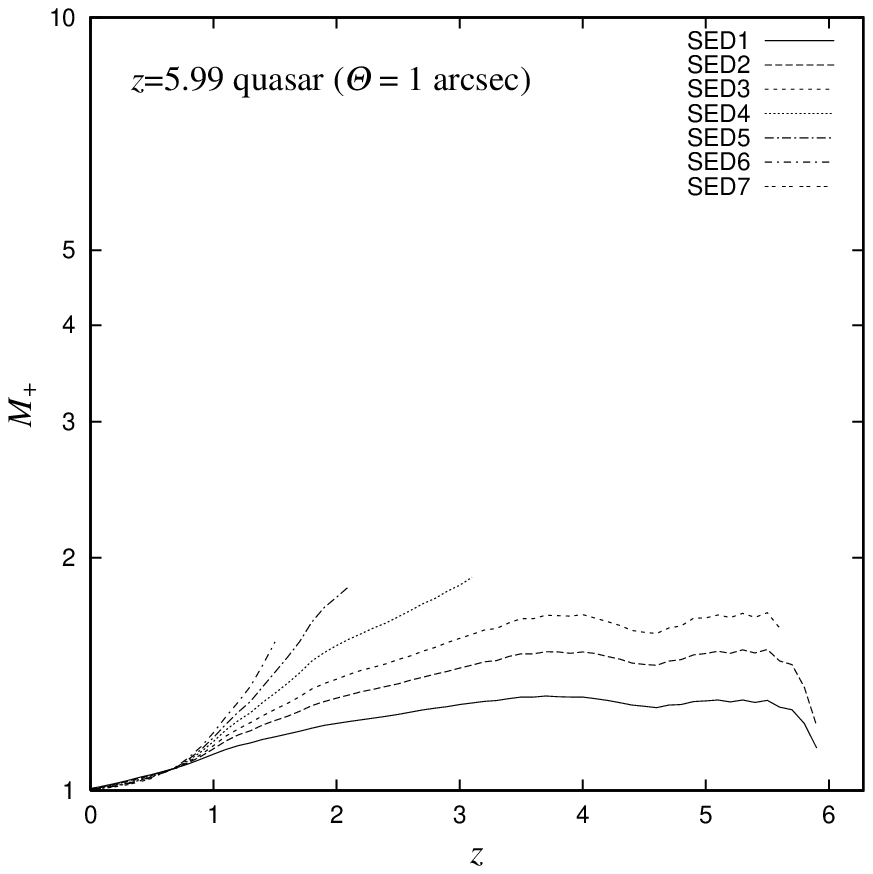}
\end{tabular}
\end{center}
\caption{
Distribution of the maximum magnification factor by the lens galaxy
with $> m_{\rm lim}$ as a function of the SED and the redshift.
}
\label{fig:fig4}
\end{figure}

\begin{table}
\caption{Journal of imaging observations.}\label{tab:tab1}
\begin{center}
\begin{tabular}{cccccc}
\hline
\hline
Quasar &
Obs. date (UT) &
Total exp. (s) &
$m_{\rm lim}$(AB)$^*$ &
$m_{\rm quasar}$(AB) &
FWHM$_{\rm star}^{\dag}$ ($^{\prime \prime}$) \\
\hline
SDSSp J103027.10+052455.0 & 2002 April 14 &  900 & 25.4 & 23.23$\pm$0.07 & 1.2 \\
SDSSp J130608.26+035626.3 & 2002 April 14 & 1200 & 25.8 & 21.95$\pm$0.03 & 1.8 \\
\hline
\end{tabular}
\end{center}

$^*$ The limiting magnitude (3$\sigma$) with a
2$^{\prime\prime}$ aperture.\\
$^{\dag}$ The full width at half maximum of stellar
objects in the final image.\\
\end{table}

\end{document}